\documentclass[twocolumn,traditabstract]{aa}
\usepackage{amsmath}
\usepackage{graphicx}
\usepackage{txfonts}
\usepackage{natbib}
\usepackage{upgreek}

\newcommand{\hpatrick}[1]{}
\newcommand{\juan}[1]{{#1}}

\newcommand{\commentproof}[1]{{\bf \color{green}#1}}
\renewcommand{\commentproof}[1]{} 

\def\setsymbol#1#2{\expandafter\def\csname #1\endcsname{#2}}
\def\getsymbol#1{\csname #1\endcsname}

\def\all2013resultspapers{\nocite{planck2013-p01, planck2013-p02, planck2013-p02a, planck2013-p02d, planck2013-p02b, planck2013-p03, planck2013-p03c, planck2013-p03f, planck2013-p03d, planck2013-p03e, planck2013-p01a, planck2013-p06, planck2013-p03a, planck2013-pip88, planck2013-p08, planck2013-p11, planck2013-p12, planck2013-p13, planck2013-p14, planck2013-p15, planck2013-p05b, planck2013-p17, planck2013-p09, planck2013-p09a, planck2013-p20, planck2013-p19, planck2013-pipaberration, planck2013-p05, planck2013-p05a, planck2013-pip56, planck2013-p06b, planck2013-p01a}}

\newbox\tablebox    \newdimen\tablewidth
\def\leaderfil{\leaders\hbox to 5pt{\hss.\hss}\hfil}
\def\tablenote#1 #2\par{\begingroup \parindent=0.8em
    \abovedisplayshortskip=0pt\belowdisplayshortskip=0pt
    \noindent
    $$\hss\vbox{\hsize\tablewidth \hangindent=\parindent \hangafter=1 \noindent
    \hbox to \parindent{$^#1$\hss}\strut#2\strut\par}\hss$$
    \endgroup}

\def\L2{\ifmmode L_2\else $L_2$\fi}

\def\DeltaT{\ifmmode \Delta T\else $\Delta T$\fi}
\def\deltat{\ifmmode \Delta t\else $\Delta t$\fi}
\def\fknee{\ifmmode f_{\rm knee}\else $f_{\rm knee}$\fi}
\def\Fmax{\ifmmode F_{\rm max}\else $F_{\rm max}$\fi}
\def\solar{\ifmmode{\rm M}_{\mathord\odot}\else${\rm M}_{\mathord\odot}$\fi}
\def\Msolar{\ifmmode{\rm M}_{\mathord\odot}\else${\rm M}_{\mathord\odot}$\fi}
\def\Lsolar{\ifmmode{\rm L}_{\mathord\odot}\else${\rm L}_{\mathord\odot}$\fi}
\def\inv{\ifmmode^{-1}\else$^{-1}$\fi}
\def\mo{\ifmmode^{-1}\else$^{-1}$\fi}
\def\sup#1{\ifmmode ^{\rm #1}\else $^{\rm #1}$\fi}
\def\expo#1{\ifmmode \times 10^{#1}\else $\times 10^{#1}$\fi}
\def\,{\thinspace}
\def\lsim{\mathrel{\raise .4ex\hbox{\rlap{$<$}\lower 1.2ex\hbox{$\sim$}}}}
\def\gsim{\mathrel{\raise .4ex\hbox{\rlap{$>$}\lower 1.2ex\hbox{$\sim$}}}}

\def\simprop{\mathrel{\raise .4ex\hbox{\rlap{$\propto$}\lower 1.2ex\hbox{$\sim$}}}}
\def\deg{\ifmmode^\circ\else$^\circ$\fi}
\def\pdeg{\ifmmode $\setbox0=\hbox{$^{\circ}$}\rlap{\hskip.11\wd0 .}$^{\circ}
          \else \setbox0=\hbox{$^{\circ}$}\rlap{\hskip.11\wd0 .}$^{\circ}$\fi}
\def\arcs{\ifmmode {^{\scriptstyle\prime\prime}}
          \else $^{\scriptstyle\prime\prime}$\fi}
\def\arcm{\ifmmode {^{\scriptstyle\prime}}
          \else $^{\scriptstyle\prime}$\fi}
\newdimen\sa  \newdimen\sb
\def\parcs{\sa=.07em \sb=.03em
     \ifmmode \hbox{\rlap{.}}^{\scriptstyle\prime\kern -\sb\prime}\hbox{\kern -\sa}
     \else \rlap{.}$^{\scriptstyle\prime\kern -\sb\prime}$\kern -\sa\fi}
\def\parcm{\sa=.08em \sb=.03em
     \ifmmode \hbox{\rlap{.}\kern\sa}^{\scriptstyle\prime}\hbox{\kern-\sb}
     \else \rlap{.}\kern\sa$^{\scriptstyle\prime}$\kern-\sb\fi}
\def\ra[#1 #2 #3.#4]{#1\sup{h}#2\sup{m}#3\sup{s}\llap.#4}
\def\dec[#1 #2 #3.#4]{#1\deg#2\arcm#3\arcs\llap.#4}
\def\deco[#1 #2 #3]{#1\deg#2\arcm#3\arcs}
\def\rra[#1 #2]{#1\sup{h}#2\sup{m}}

\def\dots{\relax\ifmmode \ldots\else $\ldots$\fi}
\def\WHzsr{\ifmmode $W\,Hz\mo\,sr\mo$\else W\,Hz\mo\,sr\mo\fi}
\def\mHz{\ifmmode $\,mHz$\else \,mHz\fi}
\def\GHz{\ifmmode $\,GHz$\else \,GHz\fi}
\def\mKs{\ifmmode $\,mK\,s$^{1/2}\else \,mK\,s$^{1/2}$\fi}
\def\muKs{\ifmmode \,\mu$K\,s$^{1/2}\else \,$\mu$K\,s$^{1/2}$\fi}
\def\muKRJs{\ifmmode \,\mu$K$_{\rm RJ}$\,s$^{1/2}\else \,$\mu$K$_{\rm RJ}$\,s$^{1/2}$\fi}
\def\muKHz{\ifmmode \,\mu$K\,Hz$^{-1/2}\else \,$\mu$K\,Hz$^{-1/2}$\fi}
\def\MJysr{\ifmmode \,$MJy\,sr\mo$\else \,MJy\,sr\mo\fi}
\def\MJysrmK{\ifmmode \,$MJy\,sr\mo$\,mK$_{\rm CMB}\mo\else \,MJy\,sr\mo\,mK$_{\rm CMB}\mo$\fi}
\def\microns{\ifmmode \,\mu$m$\else \,$\mu$m\fi}

\def\muK{\ifmmode \,\mu$K$\else \,$\mu$\hbox{K}\fi}
\def\microK{\ifmmode \,\mu$K$\else \,$\mu$\hbox{K}\fi}
\def\muW{\ifmmode \,\mu$W$\else \,$\mu$\hbox{W}\fi}
\def\kms{\ifmmode $\,km\,s$^{-1}\else \,km\,s$^{-1}$\fi}
\def\kmsMpc{\ifmmode $\,\kms\,Mpc\mo$\else \,\kms\,Mpc\mo\fi}
\providecommand{\sorthelp}[1]{}

\providecommand{\sorthelp}[1]{}
\newcommand{\kps}{km\,s$^{-1}$}

\begin{document}


\title{The filamentary structures in the CO emission toward the Milky Way disk}
\titlerunning{The filamentary structures in the CO emission toward the Milky Way disk}
    \author{
        J.~D.~Soler$^{1}$\thanks{Corresponding author, \email{soler@mpia.de}},
        H.~Beuther$^{1}$,
        J.~Syed$^{1}$,
        Y.~Wang$^{1}$, 
        Th.~Henning$^{1}$,
        S.\,C.\,O.~Glover$^{2}$,
        R.\,S.~Klessen$^{2}$
        M.\,C.~Sormani$^{2}$,
        M.~Heyer$^{3}$,
        R.\,J.~Smith$^{4}$,
        J.\,S.~Urquhart$^{5}$
        J.~Yang$^{6}$, 
        Y.~Su$^{6}$,  
        X.~Zhou$^{6}$
} 
\institute{
1. Max-Planck-Institute for Astronomy, K\"{o}nigstuhl 17, 69117, Heidelberg, Germany. \\
2. Universit\"{a}t Heidelberg, Zentrum f\"{u}r Astronomie, Institut f\"{u}r Theoretische Astrophysik, Albert-Ueberle-Str. 2, 69120, Heidelberg, Germany.\\
3. Department of Astronomy, University of Massachusetts, Amherst, MA 01003-9305, USA.\\
4. Jodrell Bank Centre for Astrophysics, School of Physics and Astronomy, University of Manchester, Oxford Road, Manchester M13 9PL, UK.\\
5. Centre for Astrophysics and Planetary Science, University of Kent, Canterbury CT2 7NH, UK.\\
6. Purple Mountain Observatory, Chinese Academy of Sciences, 10 Yuanhua Road, Qixia District, Nanjing 210033, China.
}
\authorrunning{Soler,\,J.D. et al.}

\date{Received 17MAY2021 / Accepted 16JUN2021}

\abstract{
We present a statistical study of the filamentary structure orientation in the CO emission observations obtained in the Milky Way Imaging Scroll Painting (MWISP) survey in the range 25\pdeg8\,$<$\,$l$\,$<$\,49\pdeg7, $|b|$\,$\leq$\,1\pdeg25, and $-100$\,$<$\,$v_{\rm LSR}$\,$<$\,135\,\kps.
We found that most of the filamentary structures in the $^{12}$CO and $^{13}$CO emission do not show a global preferential orientation either parallel or perpendicular to the Galactic plane.
However, we found ranges in Galactic longitude and radial velocity where the $^{12}$CO and $^{13}$CO filamentary structures are parallel to the Galactic plane.
These preferential orientations are different from those found for the H{\sc i} emission.
We consider this an indication that the molecular structures do not simply inherit these properties from parental atomic clouds.
Instead, they are shaped by local physical conditions, such as stellar feedback, magnetic fields, and Galactic spiral shocks.
}

\maketitle

\section{Introduction}\label{section:introduction}

The cold molecular gas is the main reservoir from which stars are formed in the Milky Way and in similar spiral galaxies \citep[see, for example,][]{kennicutt2012,dobbs2014,molinari2014}.
For that reason, the study of the molecular gas structure is crucial for any understanding of the star formation process \citep[see][for a review]{ballesteros-paredes2020,chevance2020}.
In this letter, which is a follow-up to the atomic hydrogen (H{\sc i}) analysis presented in \cite{soler2020}, we report the study of the filamentary structures in the emission from carbon monoxide (CO), one of the primary tracers of molecular gas.

The main focus of CO emission studies has traditionally been molecular clouds (MCs), which correspond to objects of between a few and hundreds of parsecs in size as defined by their spectral and spatial association \citep[see, for example,][]{larson1981,solomon1987,heyer2009,miville-deschenes2017}.
The estimated MC properties, such as velocity line width, size, virial mass, and luminosity, are used to evaluate the dynamical state of the molecular gas and the interplay of turbulence and self-gravity in the molecular phase of the interstellar medium (ISM).
More recently, special consideration has been given to elongated structures called giant molecular filaments (GMFs), which range between 10 and 300\,pc in length and 1 to 40\,pc in width \citep[][]{goodman2014,ragan2014,wang2015,zucker2015,abreu-vicente2017,wang2020gmf}.
Such GMFs can potentially trace the global spiral structure, such as arms and spurs, and large concentrations of molecular gas.

This letter is devoted to studying two statistical properties of the CO emission without reference to particular objects, such as MCs or GMFs.
The first property is the presence of elongated structures in the emission maps, which we characterize using the Hessian matrix, a tool broadly used in the study of H{\sc i} and other ISM tracers \citep[see, for example,][]{polychroni2013,planck2014-XXXII,schisano2014,kalberla2016}.
The second property is the orientation of the elongated structures, which we evaluate by using the tools of circular statistics \citep[see, for example,][]{batschelet1981}.

The main goal of this study is the comparison with the results of the same type of analysis applied to the HI observations in The H{\sc i}/OH/Recombination (THOR) line  survey of the inner Milky Way \citep[][]{beuther2016,wang2020hi}.
\cite{soler2020} report that most of the filamentary structures in the H{\sc i} emission are aligned with the Galactic plane.
However, they also find ranges in Galactic longitude and radial velocity where the H{\sc i} filamentary structures are primarily perpendicular to the Galactic plane.
\juan{These are located: ({\it i}) around the tangent point of the Scutum spiral arm and the terminal velocities of the molecular ring near $l$\,$\approx$\,28$\deg$ and $v_{\rm LSR}$\,$\approx$\,100\,km\,s$^{-1}$, ({\it ii}) toward $l$\,$\approx$\,35$\deg$ and $v_{\rm LSR}$\,$\approx$\,50\,km\,s$^{-1}$, ({\it iii}) around the Riegel-Crutcher cloud, and ({\it iv}) toward the positive and negative terminal velocities.}
A comparison with numerical simulations indicates that the prevalence of horizontal filamentary structures is most likely due to large-scale Galactic dynamics.
The vertical structures may arise from the combined effect of supernova (SN) feedback and strong magnetic fields \citep{hennebelle2018,smith2020}.
To test if the trends found in the H{\sc i} emission are also present in the molecular gas, we applied the Hessian matrix analysis to the $^{12}$CO, $^{13}$CO, and C$^{18}$O observations from the Milky Way Imaging Scroll Painting (MWISP) survey \citep[][]{su2019}.

\section{Data and methodology}\label{section:data}

\subsection{The Milky Way Imaging Scroll Painting (MWISP)}\label{section:MWISPdata}

The MWISP project is a high-sensitivity survey of the northern Galactic plane observed with the Purple Mountain Observatory 13.7 m telescope \citep{su2018,su2019}.
In this work we used the published data from the survey in the region within 25\pdeg8$<$\,$l$\,$<$\,49\pdeg7 and $|b|$\,$<$\,5\pdeg2, although we exclusively considered the range $|b|$\,$<$\,1\pdeg25 to match the range of the H{\sc i} data in the \cite{soler2020} analysis.

The $^{12}$CO($J$=1--0), $^{13}$CO($J$=1--0), and C$^{18}$O ($J$=1--0) lines are simultaneously observed by the nine-beam Superconducting Spectroscopic ARray (SSAR) receiving system \citep{shan2012}.
The full-width half maximum (FWHM) of the observations is 49\arcs\ at the $^{12}$CO frequency and 51\arcs\ at the $^{13}$CO and C$^{18}$O frequencies, respectively.
A total bandwidth of 1\,GHz with 16,384 channels provides a velocity coverage of 260\,\kps\ and a spectral resolution of 61\,kHz, equivalent to velocity separations of about 0.16\,\kps\ for $^{12}$CO and 0.17\,\kps\ for $^{13}$CO and C$^{18}$O.
This velocity coverage includes the portion of negative radial velocities ($v_{\rm LSR}$\,$<$\,0\,\kps), which is excluded in other high-resolution CO surveys of the Galactic plane, such as the Boston University-Five College Radio Astronomy Observatory Galactic Ring Survey \citep[GRS;][]{jackson2006} and the FOREST Unbiased Galactic plane Imaging survey with the Nobeyama 45 m telescope \citep[FUGIN;][]{unemoto2017}.
In this letter we have used these data, which we smoothed and re-gridded to the THOR H{\sc i} 1.5\,\kps spectral resolution by using the tools in the {\tt spectral-cube} package in {\tt astropy} \citep[][]{astropy:2018}
The results obtained at the native spectral resolution are presented in Appendix~\ref{app:hessian}.

The typical root-mean-square (RMS) noise levels are about 0.5\,K for $^{12}$CO and 0.3\,K for $^{13}$CO and C$^{18}$O, respectively.
\juan{The final data products are position-position-velocity (PPV) cubes constructed from a mosaic with a spatial grid spacing of 30\arcsec.}
We projected these data into the 10\arcsec\,$\times$10\arcsec\ spatial grid of the THOR H{\sc i} data by using the {\tt reproject} package, also from {\tt astropy}.

\subsection{The Hessian matrix method}\label{sec:hessian}

We applied the method described in \cite{soler2020} to the MWISP CO observations.
For each position of an intensity map corresponding to $v_{\rm LSR}$\,$=$\,$v$, $I(l,b,v)$, we estimated the derivatives with respect to the local coordinates $(x,y)$ and built the Hessian matrix,
\begin{equation}
\mathbf{H}(x,y) \equiv \begin{bmatrix} 
H_{xx} & H_{xy} \\
H_{yx} & H_{yy} 
\end{bmatrix},
\end{equation}
where $H_{xx}$\,$\equiv$\,$\partial^{2} I/\partial x^{2}$, $H_{xy}$\,$\equiv$\,$\partial^{2} I/\partial x \partial y$, $H_{yx}$\,$\equiv$\,$\partial^{2} I/\partial y \partial x$, $H_{yy}$\,$\equiv$\,$\partial^{2} I/\partial y^{2}$, and $x$ and $y$ are related to the Galactic coordinates $(l, b)$ as $x$\,$\equiv$\,$l\cos b$ and $y$\,$\equiv$\,$b$, so that the $x$-axis is parallel to the Galactic plane.
We obtained the partial spatial derivatives by convolving $I(l,b,v)$ with the second derivatives of a two-dimensional Gaussian function.
In practice, we used the {\tt gaussian\_filter} function in the open-source software package {\tt SciPy}.
To match the results of \cite{soler2020}, we selected a derivative kernel with a 120\arcsec\ FWHM.
The results obtained with different derivative kernel sizes and $\Delta v$ selections are presented in Appendix~\ref{app:hessian}.

The two eigenvalues ($\lambda_{\pm}$) of the Hessian matrix were found by solving the characteristic equation,
\begin{equation}\label{eq:lambda}
\lambda_{\pm} = \frac{(H_{xx}+H_{yy}) \pm \sqrt{(H_{xx}-H_{yy})^{2}+4H_{xy}H_{yx}}}{2}.
\end{equation}
Both eigenvalues define the local curvature of the intensity map.
In particular, the minimum eigenvalue ($\lambda_{-}$) highlights filamentary structures or ridges, as detailed in \cite{planck2014-XXXII}.
The eigenvector corresponding to $\lambda_{-}$ defines the orientation of intensity ridges with respect to the Galactic plane, which is characterized by the angle
\begin{equation}\label{eq:theta}
\theta = \frac{1}{2}\arctan\left[\frac{H_{xy}+H_{yx}}{H_{xx}-H_{yy}}\right].
\end{equation}
We estimated an angle $\theta_{ij}$ for each of the $m$\,$\times$\,$n$ pixels in a velocity channel map, where the indices $i$ and $j$ run over the pixels along the $x$- and $y$-axis, respectively.
This angle, however, is only meaningful in regions of the map that are rated as filamentary according to selection criteria based on the values of $\lambda_{-}$ and on the noise properties of the data.

We conducted the Hessian analysis in 2$\deg$\,$\times$\,2$\deg$ tiles and across velocity channels, both at their native resolution and at $\Delta v$\,$=$\,1.5\,\kps\ for the comparison with THOR H{\sc i} data.
In each tile, we selected the filamentary structures based on the criterion $\lambda_{-}$\,$<$\,$0$.
Additionally, we selected regions where $I(l,b,v)$\,$>$\,3$\sigma_{I}$, where $\sigma_{I}$ corresponds to the RMS noise presented in Sect.~\ref{section:MWISPdata}.
Following the method introduced in \cite{planck2014-XXXII}, we further selected the filamentary structures depending on the values of the eigenvalue $\lambda_{-}$ in noise-dominated portions of the data.
For that purpose, we estimated $\lambda_{-}$ in five velocity channels with low signal-to-noise ratios and determined the minimum value of $\lambda_{-}$ in each of them.
We used the median of these five $\lambda_{-}$ values as the threshold value, $\lambda^{C}_{-}$.
\juan{We employed the median to reduce the effect of outliers, but in general the values of $\lambda_{-}$ in the noise-dominated channels are similar and this selection does not imply any loss of generality.}
We exclusively considered regions of each velocity channel map where $\lambda_{-}$\,$<$\,$\lambda^{C}_{-}$, which corresponds to the selection of filamentary structures with curvatures in $I(l,b,v)$ larger than those present in the noise-dominated channels.

Once the filamentary structures were selected, we used the angles derived from Eq.~\ref{eq:theta} to study their orientation with respect to the Galactic plane.
\juan{For a systematic evaluation of the preferential orientation, we applied the projected Rayleigh statistic ($V$) \citep[see, for example,][]{batschelet1981}, which is a test to determine whether the distribution of angles is nonuniform and peaked at a particular angle.}
This test is equivalent to the modified Rayleigh test for uniformity proposed by \cite{durandANDgreenwood1958} for the specific directions of interest $\theta$\,$=$\,$0\deg$ and $90\deg$ \citep{jow2018}, such that $V$\,$>$\,0 or $V$\,$<$\,0 correspond to preferential orientations parallel or perpendicular to the Galactic plane, respectively.
It is defined as
\begin{equation}\label{eq:myprs}
V = \frac{\sum^{n,m}_{ij}w_{ij}\cos(2\theta_{ij})}{\sqrt{\sum^{n,m}_{ij}w_{ij}/2}},
\end{equation}
where the indices $i$ and $j$ run over the pixel locations in the two spatial dimensions $(l,b)$ for a given velocity channel and $w_{ij}$ is the statistical weight of each angle $\theta_{ij}$.

The values of $V$ lead to significance only if there is sufficient clustering around the orientations $\theta$\,$=$\,$0\deg$ and $ 90\deg$.
The null hypothesis implied in $V$ is that the angle distribution is uniform or centered on a different orientation angle.
In the particular case of independent and uniformly distributed angles, and for a large number of samples, values of $V$\,$\approx$\,$1.64$ and $2.57$ correspond to the rejection of the null hypothesis with a probability of 5\% and 0.5\%, respectively \citep{batschelet1972}.
A value of $V$\,$\approx$\,2.87 is roughly equivalent to a 3$\sigma$ confidence interval.
\juan{In our application, we accounted for the spatial correlations introduced by the telescope beam by choosing $w_{ij}$\,$=$\,$(\Delta x/D)^{2}$, where $\Delta x$ is the pixel size and $D$ is the diameter of the derivative kernel that we selected to calculate the gradients.}
We note, however, that the correlation across scales in the ISM makes it very difficult to estimate the absolute statistical significance of $V$.
Further details on the significance of $V$ are presented in Appendix~\ref{app:circstats}.

\section{Results}\label{section:results}

\begin{figure}[ht!]
\centerline{\includegraphics[width=0.5\textwidth,angle=0,origin=c]{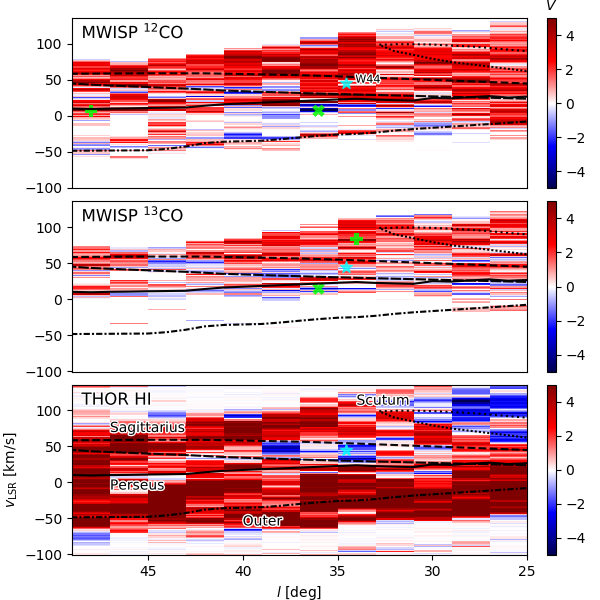}}
\caption{Projected Rayleigh statistic ($V$) corresponding to the orientation of the filamentary structures identified in the 2$\deg$\,$\times$\,2$\deg$ tiles and across velocity channels in the THOR H{\sc i} and MWISP $^{12}$CO and $^{13}$CO observations.
The overlaid curves correspond to selected spiral arms from the model presented in \cite{reid2016}.
The plus sign ($+$) and the cross ($\times$) mark the tiles with the highest and the lowest $V$, respectively, whose maps are shown for reference in Fig.~\ref{fig:maxPRStile12}.
The star symbols mark the positions of the SN remnant W44.
}
\label{fig:lvdiagTHORandMWISP_PRS}
\end{figure}

The outcome of the Hessian analysis of the MWISP data is summarized in Fig.~\ref{fig:lvdiagTHORandMWISP_PRS}.
The amount of emission that is classified as filaments varies throughout the $l$ and $v_{\rm LSR}$, but it corresponds to up to 60\,\% of the area in some tiles, as further illustrated in Appendix~\ref{app:hessian}.
The first compelling result of this analysis is the fact that the majority of the tiles with significant CO detections show $V$\,$>$\,0, which suggests a preference for the filamentary structures to be parallel to the Galactic plane. 
However, only 26\% and 9\% of the valid $^{12}$CO and $^{13}$CO tiles show $V$\,$>$\,2.87, which is the 3$\sigma$ threshold that indicates an unequivocal preference for the structures to be aligned with the Galactic plane.
Fewer than 1\% of the tiles show $V$\,$<$\,$-$2.87, which is the corresponding threshold that indicates the preference for the filamentary structures to be perpendicular to the Galactic plane.
According to that statistical significance criterion, the large majority of the filamentary structures have no preference to be either parallel or perpendicular to the Galactic plane, with $|V|$\,$<$\,2.87 in roughly 73\% and 90\% of the valid $^{12}$CO and $^{13}$CO tiles, respectively.
The C$^{18}$O emission is in general less extended, and only 23 of the 1710 tiles with significant detections show $|V|$\,$>$\,2.87, which is insufficient for a global comparison.
These percentages do not change significantly when a higher detection threshold in the CO emission is selected, as illustrated in Appendix~\ref{app:hessian}.
\juan{There are no significant preferential orientations in the integrated emission (moment-zero) maps, as also shown in Appendix~\ref{app:hessian}.}

Most of the $V$\,$>$\,0 values appear in the $v_{\rm LSR}$\,$>$\,0\,\kps\ range, and there is no apparent correlation between the filament orientation and the velocities associated with spiral arms.
For $v_{\rm LSR}$\,$<$\,0\,\kps, the trend in $V$ is less homogeneous and there is a variety of positive and negative $V$ values. 
Most of the tiles with negative values of $V$, which corresponds to filamentary structures perpendicular to the Galactic plane, are found in the velocity range between the Perseus and the Outer spiral arms.

\begin{figure*}[ht!]
\centerline{\includegraphics[width=0.99\textwidth,angle=0,origin=c]{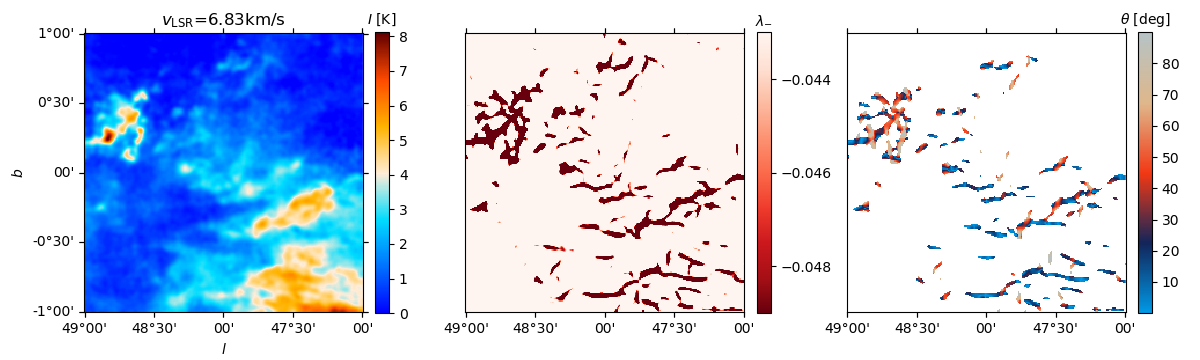}}
\vspace{-0.4cm}
\centerline{\includegraphics[width=0.99\textwidth,angle=0,origin=c]{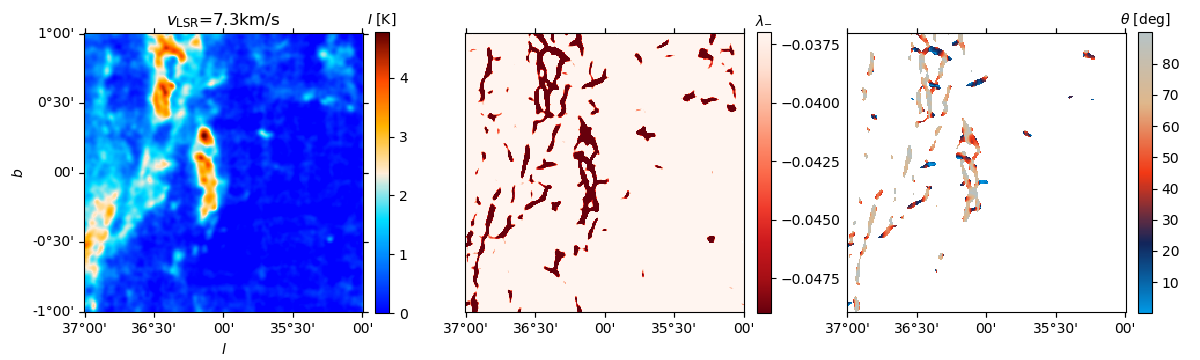}}
\caption{
Results of the Hessian analysis in the 2$\deg$\,$\times$\,2$\deg$ tiles of the MWISP $^{12}$CO observations with the maximum (top) and minimum (bottom) values of $V$, which can be interpreted as a tile dominated by horizontal structures and a tile dominated by vertical structures, respectively.
{\it Left}. Intensity map.
{\it Center}. Map of the eigenvalue of the Hessian matrix identified as $\lambda_{-}$ in Eq.~\eqref{eq:lambda}, which is used to characterize the filamentary structures in the intensity map.
{\it Right}. Map of the orientation angles evaluated using Eq.~\eqref{eq:theta} after the selections by $I$ and $\lambda_{-}$ introduced in Sect.~\ref{sec:hessian}.
}
\label{fig:maxPRStile12}
\end{figure*}

Figure~\ref{fig:maxPRStile12} shows the maps of the tiles with the extreme positive and negative $V$ values in $^{12}$CO, which correspond to the $l$ and $v_{\rm LSR}$ indicated in the top panel of Fig.~\ref{fig:lvdiagTHORandMWISP_PRS}.
The panel showing the maximum value of $V$, which corresponds to mostly horizontal structures, shows that the trend does not correspond to a monolithic filament parallel to the Galactic plane, but rather to a set of structures from which it is difficult to guess the position of the Galactic disk.
The panel showing the minimum value of $V$, which corresponds to mostly vertical structures, shows structures that are reminiscent of the intermediate-velocity clouds identified by \cite{soler2020} at the terminal velocities in H{\sc i} emission.

\begin{figure*}[ht!]
\centerline{
\includegraphics[width=0.99\textwidth,angle=0,origin=c]{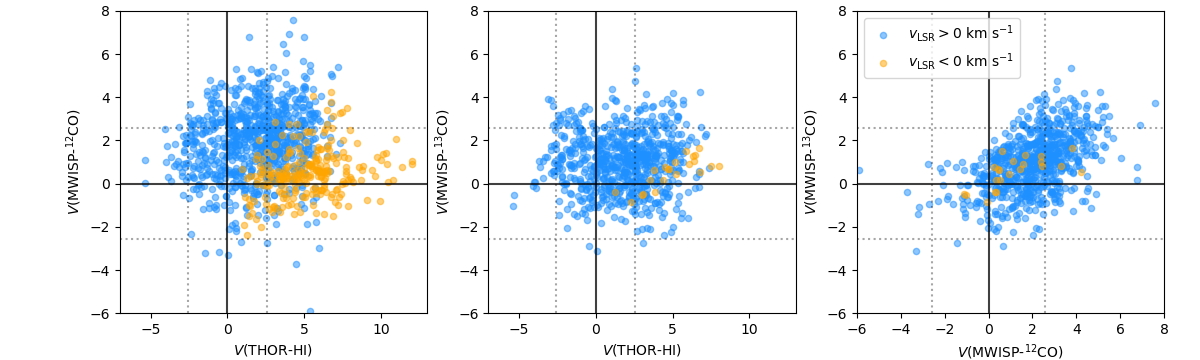}
}
\caption{Comparison between the values of the projected Rayleigh statistic ($V$) in the THOR H{\sc i} and the MWISP CO observations.
The two colors correspond to the positive and negative radial velocities.
The dotted lines mark the 3$\sigma$ significance thresholds in $V$.
}
\label{fig:lvdiagramMultiTracer}
\end{figure*}

Figure~\ref{fig:lvdiagTHORandMWISP_PRS} shows that the tiles with the most prominent $V$\,$<$\,0 value in H{\sc i} do not have a similar behavior in $^{12}$CO and $^{13}$CO. 
This is further illustrated in Fig.~\ref{fig:lvdiagramMultiTracer}, which shows that the relatively large $V$ values found for the H{\sc i} emission, in the $l$ and $v_{\rm LSR}$ ranges mention in Sect.~\ref{section:introduction}, are not found in the $^{12}$CO or $^{13}$CO emission.
The right-hand-side panel of Fig.~\ref{fig:lvdiagramMultiTracer} also indicates that, in general, the largest values of $V$ are found at $v_{\rm LSR}$\,$<$\,0\,\kps\ for H{\sc i} and $v_{\rm LSR}$\,$>$\,0\,\kps\ for $^{12}$CO.
\juan{This trend is also present in $^{13}$CO, but with fewer valid tiles in the $v_{\rm LSR}$\,$<$\,0\,\kps\ range.}
The comparison of the $V$ values in $^{12}$CO and $^{13}$CO, presented in the right-hand-side panel of Fig.~\ref{fig:lvdiagramMultiTracer}, indicates that there is a general agreement in the orientation of the structures sampled by the two tracers, but the dominant filament orientation in one of them is not indicative of the  orientation in the other.

\section{Discussion}

Our general results indicate that the alignment of the filamentary structures in the $^{12}$CO and $^{13}$CO emission with the Galactic plane is not very significant in the 25\pdeg8\,$<$\,$l$\,$<$\,49\pdeg7 range.
This observation is an indication of a potential selection effect in the general trends found for the GMFs, which appear aligned with the Galactic plane \citep[see][for a review]{zucker2018}.
This selection effect may potentially be related to GMF identification via visual inspection of the extinction maps, which is biased toward the densest and most conspicuous structures.
Furthermore, this indicates that the general Galactic-plane alignment of the filamentary structures identified by \cite{li2016} in the dust thermal emission toward the inner Galaxy, $-$60\pdeg0\,$<$\,$l$\,$<$\,60\pdeg0, may be the product of the line-of-sight integration or a selection effect introduced by that particular tracer.

The results presented in Fig.~\ref{fig:lvdiagTHORandMWISP_PRS} show significant differences between the CO filamentary structure orientation at $v_{\rm LSR}$\,$<$\,0 and at $v_{\rm LSR}$\,$>$\,0\,\kps. 
If we consider kinematic distances, positive $v_{\rm LSR}$ values in this $l$ range correspond to objects within roughly 10 \,kpc of the Sun and negative values to objects that are farther away.
Thus, it is plausible that the observed trends correspond to the flaring of the disk in the outer Galaxy \citep{lozinskaya1963,levine2006}. 
However, this should also affect the H{\sc i} filaments, which is not what is observed.
Alternatively, it may be the result of the mapping of many of the density structures in position-position-position (PPP) space being crammed into the same velocity channel in PPV, an effect called velocity crowding \citep[see, for example,][]{beaumont2013}.
This effect is more acute at $v_{\rm LSR}$\,$>$\,0\,\kps\ due to the overlap of at least two distance ranges in the same radial velocity, even if we assume purely circular motions.

Another compelling result of our analysis is the variation in filament orientations along the $l$ and $v_{\rm LSR}$ corresponding to the Perseus and Outer spiral arms.
If the filamentary structures are bone-like feature of the spiral arms, as could be the case for some GMFs \citep[][]{goodman2014}, these variations are potential indications that segments of the spiral arm are being disrupted \citep[][]{tchernyshyov2018}.
While one could expect alignment of the GMF at the spiral shock, the observations may correspond to features farther downstream, which could be torqued into misalignments by the action of turbulence in the spiral arm \citep[see, for example,][]{kim2006}.
Testing whether or not these variations in filament orientation are indicative of the dynamic spiral arm structure calls for further studies based on numerical simulations and synthetic observations. 

The scatter plots presented in Fig.~\ref{fig:lvdiagramMultiTracer} show a general lack of correlation in the orientation of the H{\sc i} and CO filamentary structures.
There can be several reasons for this dissimilar behavior.
First, the line widths of the $^{13}$CO emission are narrower than those of the H{\sc i,} and it is possible that we are washing away part of the orientation of the filaments by projecting both data sets into the same spectral grid.
This effect, however, is not dominant in the orientation of the structures, as expected from the CO line width distributions \citep[see, for example][]{riener2020} and confirmed by the results of the analysis at the native velocity resolution of 0.16\,\kps\, presented in Appendix~\ref{app:hessian}.
Second, the much larger filling factor of the H{\sc i} makes it unlikely that most of its structure is related to that of the less-filling molecular gas, even if there is a considerable amount of diffuse CO gas \citep{roman-duval2016}.
Thus, although there is a morphological correlation of the H{\sc i} and the $^{13}$CO, as quantified in \cite{soler2019}, the filamentary structure in the H{\sc i} emission is not necessarily related to that of the CO, simply because they are tracing different objects.
However, when evaluating comparable scales, the H{\sc i} and the CO can appear decoupled, as shown, for example, in the structures studied in \cite{beuther2020} and \cite{syed2020hisa}.

Both the H{\sc i} and CO are subject to the same large-scale gravitational potential, which establishes the Galactic plane as the main axis of symmetry.
Thus, the difference in the orientations of the structures sampled by each tracer can be assigned to more localized physical conditions, such as self-gravity, magnetic fields, or stellar feedback.
The aforementioned processes affect each of the ISM phases in a different way and depend on the Galactic environment \citep[see, for example,][]{dale2015,krumholz2019,hennebelleANDinutsuka2019}.

One clear illustration of dissimilar filament orientations in the atomic and molecular tracers is found toward W44, a prominent SN remnant whose position and central velocity are marked in Fig.~\ref{fig:lvdiagTHORandMWISP_PRS}.
While the vicinity of this region is dominated by vertical filaments in H{\sc i}, the orientation of the CO filaments remains mostly parallel to the Galactic plane.
This can be the result of the limited effect of SN feedback on disrupting the dense molecular gas, either due to the lack of clustering or to the location of the explosion \citep{hennebelle2014,walch2015,kim2017,tress2020}.

In general, the H{\sc i} is less dense and can be more readily structured by the Galactic fountain mechanism \citep{shapiro1976,bregman1980,fraternali2017,kim2018}.
This is supported by our findings around the aforementioned SN remnants and at the terminal velocities, where the vertical H{\sc i} clouds identified as the  ``extraplanar'' H{\sc i} gas falling into the disk \citep{shane1971,lockman2002} do not have a counterpart in CO.
\juan{The CO gas is denser and less prone to being structured by this mechanism, instead being shaped by the increases in density and extinction that lead to the formation of molecules \citep{reach1994,draineANDbertoldi1996,gloverANDmaclow2011}.}

The difference in the orientation in the H{\sc i} and CO filaments can also be produced by interstellar magnetic fields.
The HI gas is typically magnetically subcritical, and its structure tends to appear parallel to the magnetic field lines \citep{clark2014,planck2014-XXXII}.
A significant portion of the molecular gas traced by CO is magnetically supercritical, and its structure tends to appear perpendicular to the magnetic field lines \citep{planck2015-XXXV,fissel2019,heyer2020}.
Therefore, it is plausible that the magnetic field geometry has a more pronounced effect on the H{\sc i} filaments than on the CO ones.

The degree to which stellar feedback or magnetic fields are responsible for the observed filament orientations remains to be quantified.
But so far our results are in tension with a linear scenario in which there is a direct causal relation between H{\sc i} and CO filaments.
The formation mechanisms of the CO filaments remain to be determined, as does their relation to the atomic gas reservoir.

\section{Conclusions}\label{section:conclusions}

Our statistical study of the filamentary structures in atomic and molecular tracers indicates that, in general, the H{\sc i} and the CO structures do not show the same preferential orientation.
\juan{This result suggests that the molecular structures do not simply inherit these properties from parental atomic clouds but rather are exposed to different physical conditions that may decouple them from the preferential orientation imposed by the Galactic plane.}
Moreover, the lack of agreement indicates that physical processes, such as magnetic fields, SN feedback, and Galactic spiral shocks, significantly affect the H{\sc i} distribution and, consequently, the gas available for star formation.
However, they do not directly determine the global molecular gas structure.

The statistical analysis presented in this letter offers insights into the general distribution and the coupling of the ISM phases.
It provides a broader characterization of the data that does not correspond to a simplified model of the ISM structure but rather to the trends identified in the observations.
This analysis complements object-based theoretical studies and provides a natural way to reduce the complex interaction of scales and physical processes in the ISM.

\begin{acknowledgements}
JDS and HB acknowledge funding from the European Research Council under the Horizon 2020 Framework Program via the ERC Consolidator Grant CSF-648505.
HB and JS acknowledge support by the Deutsche Forschungsgemeinschaft (DFG, German Research Foundation) project 138713538 and SFB 881 ``The Milky Way System'', subproject B01.
RJS is funded by an STFC ERF (grant ST/N00485X/1).
This research made use of the data from the Milky Way Imaging Scroll Painting (MWISP) project.
We are grateful to all the members of the MWISP working group, particularly the staff members at PMO-13.7m telescope, for their long-term support.
MWISP was sponsored by National Key R\&D Program of China with grant 2017YFA0402701 and CAS Key Research Program of Frontier Sciences with grant QYZDJ-SSW-SLH047.
JY is supported by National Natural Science Foundation of China through grant 12041305.

We thank the anonymous referee for the thorough review, which significantly contributed to improving the quality of this paper.
JDS thanks the following people who helped with their encouragement and conversation: Robin Tre\ss, Jonathan Henshaw, Andrea Bracco, and Naomi McClure-Griffiths.
This work has been written during a moment of strain for the world and its inhabitants. 
It would not have been possible without the effort of thousands of workers facing the COVID-19 emergency around the globe.
Our deepest gratitude to all of them.

\end{acknowledgements}

\bibliographystyle{aa}
\bibliography{41327corr}

\appendix 


\section{Hessian matrix method applied to the native resolution CO data}\label{app:hessian}

\begin{figure}[ht!]
\centerline{\includegraphics[width=0.5\textwidth,angle=0,origin=c]{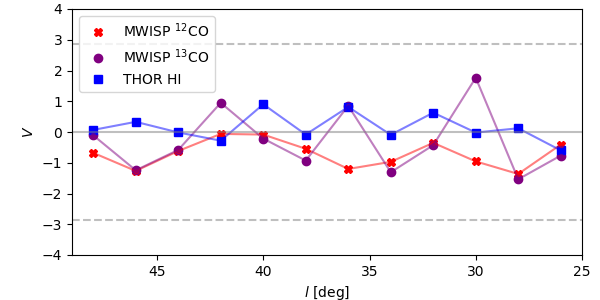}}
\caption{
Projected Rayleigh statistic ($V$) obtained for 2$\deg$\,$\times$\,2$\deg$ tiles in the integrated (moment-zero) emission maps from the THOR H{\sc i} and MWISP $^{12}$CO and $^{13}$CO observations.
The horizontal dashed lines correspond to $V$\,$=$\,$\pm$2.87, which is roughly equivalent to a 3$\sigma$ confidence interval in that statistic of an angle distribution.
}
\label{fig:lvdiagTHORandMWISPint_PRS}
\end{figure}

We present the results of the Hessian analysis applied to the integrated emission (moment-zero) maps in Fig.~\ref{fig:lvdiagTHORandMWISPint_PRS}.
The values of $|V|$\,$<$\,2.87 indicate that there is no preferential orientation in the filamentary structures in the moment-zero map.

\begin{figure*}[ht!]
\centerline{
\includegraphics[width=0.5\textwidth,angle=0,origin=c]{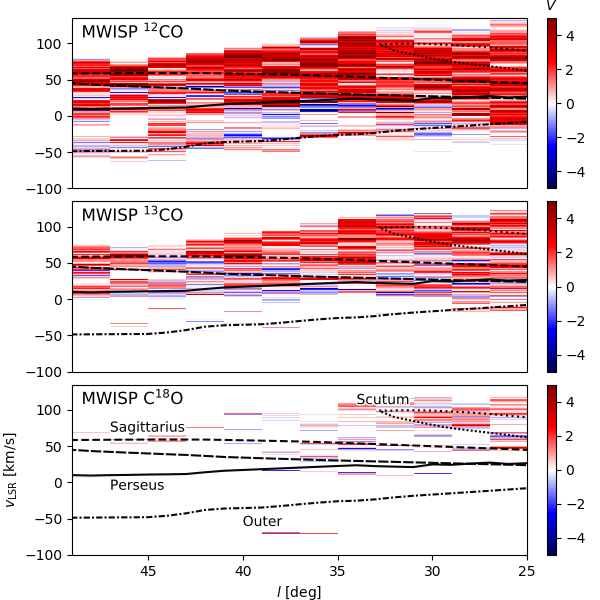}
\includegraphics[width=0.5\textwidth,angle=0,origin=c]{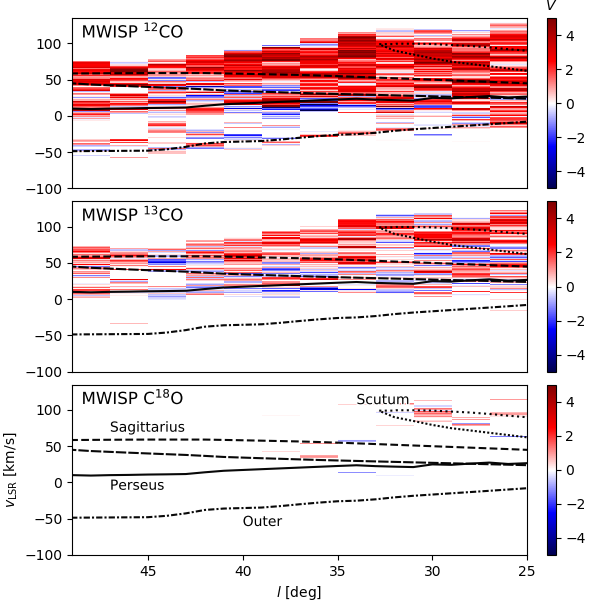}
}
\caption{Projected Rayleigh statistic ($V$) corresponding to the orientation of the filamentary structures identified in the 2$\deg$\,$\times$\,2$\deg$ tiles and across velocity channels in the MWISP CO survey with detection thresholds $I$\,$>$\,3$\sigma_{I}$ (left) and $I$\,$>$\,5$\sigma_{I}$ (right).
The overlaid curves correspond to selected spiral arms from the model presented in \cite{reid2016}.
}
\label{fig:lvdiagMWISP_PRS}
\end{figure*}

In the main body of this letter, we consider the orientation of the CO emission structures in a spectral grid that matches the THOR H{\sc i} data, $\Delta v$\,$=$\,1.5\,\kps.
For the sake of completeness, we present the results corresponding to the MWISP native resolution in Fig.~\ref{fig:lvdiagMWISP_PRS} and for two intensity thresholds, $I$\,$>$\,3$\sigma_{I}$ (left) and $I$\,$>$\,5$\sigma_{I}$ (left).

\begin{figure}[ht!]
\centerline{
\includegraphics[width=0.5\textwidth,angle=0,origin=c]{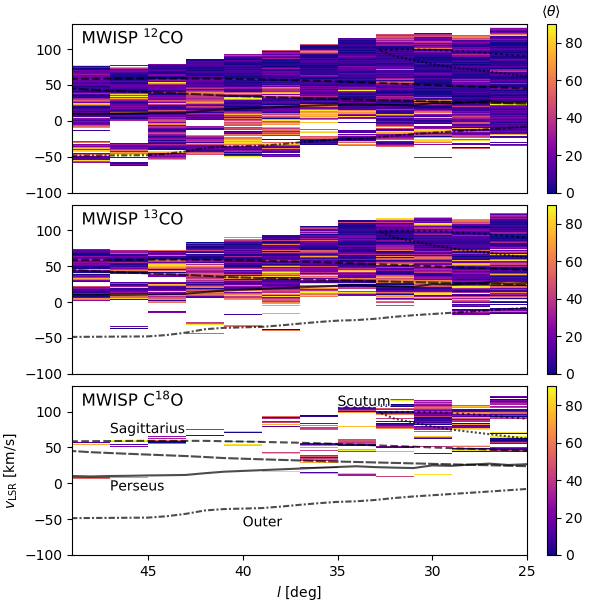}
}
\caption{
Same as Fig.~\ref{fig:lvdiagMWISP_PRS} but for the mean orientation angle ($\left<\theta\right>$).
}
\label{fig:lvdiagMWISP_meanTheta}
\end{figure}

\begin{figure}[ht!]
\centerline{\includegraphics[width=0.5\textwidth,angle=0,origin=c]{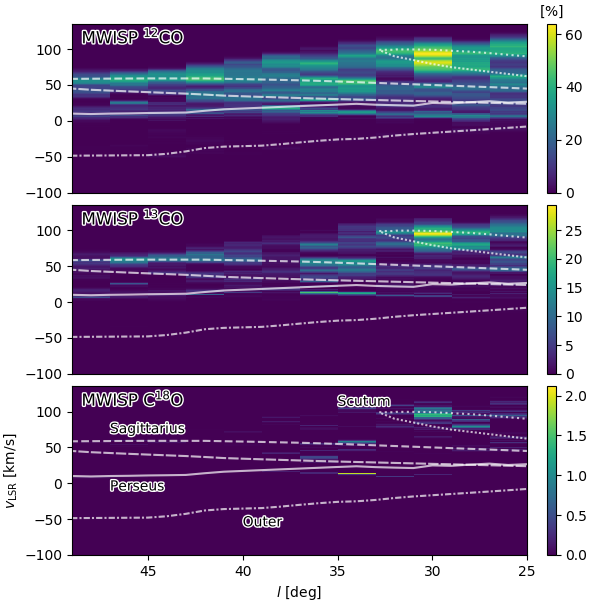}}
\caption{
Same as Fig.~\ref{fig:lvdiagMWISP_PRS} but for the percentage of the tile area that is covered by structures classified as filaments.
}
\label{fig:lvdiagMWISP_filper}
\end{figure}

We have also presented the results of the filament orientations in term of the projected Rayleigh statistic ($V$).
For the sake of completeness, here we present the result in terms of the mean orientation angle ($\left<\theta\right>$), shown in Fig.~\ref{fig:lvdiagMWISP_meanTheta}.
Additionally, we present the percentage of each 2$\deg$\,$\times$\,2$\deg$ tile covered by structures classified as filaments in Fig.~\ref{fig:lvdiagMWISP_filper}.

\section{Uncertainty in the projected Rayleigh statistic}\label{app:circstats}

The uncertainty on $V$ can be estimated by assuming that each orientation angle $\theta_{ij}$ derived from Eq.~\eqref{eq:theta} is independent and uniformly distributed, which leads to the bounded function
\begin{equation}\label{eq:s_prs}
\sigma^{2}_V = \frac{2\sum^{n}_{ij}w_{ij}\left[\cos(2\theta_{ij})\right]^{2}-V^{2}}{\sum_{ij}w_{ij}},
\end{equation}
as described in \cite{jow2018}.
In the particular case of identical statistical weights, $w_{ij}$\,$=$\,$w$, $\sigma^{2}_V$ has a maximum value of $w$.
We also estimated $\sigma^{2}_V$ by directly propagating the Monte Carlo sampling introduced in Appendix~A of \cite{soler2020}.
This method produces slightly higher values than those found using Eq.~\eqref{eq:s_prs}, most likely because it accounts for the correlation between the orientation angles in the map.


\end{document}